\documentclass[12pt]{article}
\usepackage{graphicx}

\newcommand\pubnumber{DPF2013-234}
\newcommand\pubdate{August 16, 2013}
\newcommand\preprint{FERMILAB-CONF-13-433}

\def\Fermilab-D0{Fermi National Accelerator Laboratory\\
Batavia, Illinois 60510, USA\\
for the D0 Collaboration}

\def\Title#1{\begin{center} {\Large #1 } \end{center}}
\def\Author#1{\begin{center}{ \sc #1} \end{center}}
\def\Address#1{\begin{center}{ \it #1} \end{center}}

\newcommand\pubblock{\rightline{\begin{tabular}{l} \pubnumber\\
      \pubdate\\   \preprint \end{tabular}}}
\newenvironment{Abstract}{\begin{quotation}  }{\end{quotation}}
\newenvironment{Presented}{\begin{quotation} \begin{center} 
             PRESENTED AT\end{center}\bigskip 
      \begin{center}\begin{large}}{\end{large}\end{center} \end{quotation}}
\def\Acknowledgments{\bigskip  \bigskip \begin{center} \begin{large}
             \bf ACKNOWLEDGMENTS \end{large}\end{center}}
\def\beq{\begin{equation}}
\def\eeq#1{\label{#1}\end{equation}}
\def\eeqn{\end{equation}}

\def\beqa{\begin{eqnarray}}
\def\eeqa#1{\label{#1}\end{eqnarray}}
\def\eeqan{\end{eqnarray}}

\let\bar=\overbar

\def\Dslash{\not{\hbox{\kern-4pt $D$}}}
\def\dslash{\not{\hbox{\kern-2pt $\del$}}}

\def\msb{{\bar{\ssstyle M \kern -1pt S}}}

\RequirePackage{lineno}
\begin{document}

%%% from Rick Van Kooten to add linenumbers - 9/20/2013
%\setpagewiselinenumbers
%-this works, but removed for production version \modulolinenumbers[1]
%-this works, but removed for production version \linenumbers
%%% from Rick Van Kooten to add linenumbers - 9/20/2013

\begin{titlepage}
\pubblock

\vfill
%%%%%%%%%%%%%%%%%%%%%%%%%%%%%%%%%%%%%%%%%%%%%%%%%%%%%%%%%%%%%%%%%%%%%%%%%%%%%%%%%%%%%%%%%%%%%%%%%%%%%%%%
\Title{Update of the like-sign dimuon charge asymmetry measurement from the D0 experiment}
%%%    DRAFT:  Tuesday, October 1 - 12:30 PM
%%%%%%%%%%%%%%%%%%%%%%%%%%%%%%%%%%%%%%%%%%%%%%%%%%%%%%%%%%%%%%%%%%%%%%%%%%%%%%%%%%%%%%%%%%%%%%%%%%%%%%%%
\vfill
% remove support \Author{ Peter H. Garbincius\support}
\Author{Peter H. Garbincius}
\Address{\Fermilab-D0}
\vfill

\begin{Abstract}
The D0 Collaboration has published three measurements of the CP-violating like-sign dimuon charge asymmetry in $p\overline{p}$ 
collisions at the \mbox{Fermilab} Tevatron collider. These measurements are significantly different from the standard model 
predictions. In this presentation, we discuss the status of the final measurement of this asymmetry 
and the expected improved sensitivities, using the full 10.4 fb$^{-1}$ 
data sample collected during Run II, and discuss its possible interpretations.
\end{Abstract}

\vfill
\begin{Presented}
DPF 2013\\
The Meeting of the American Physical Society\\
Division of Particles and Fields\\
Santa Cruz, California, August 13--17, 2013\\
\end{Presented}
\vfill
\end{titlepage}
\def\thefootnote{\fnsymbol{footnote}}
\setcounter{footnote}{0}

\section{Introduction:  the Mystery and the Motivation}

The single inclusive muon and like-sign dimuon charge asymmetries are defined as the ratios of the difference and sum of the rates
$a_{CP} = \{ \Gamma(\mu^+)-\Gamma(\mu^-)\}$/$ \{ \Gamma(\mu^+) + \Gamma(\mu^-) \}$ and
$A_{CP} = \{ \Gamma(\mu^+ \mu^+)-\Gamma(\mu^- \mu^-) \}$/$ \{\Gamma(\mu^+ \mu^+) +\Gamma(\mu^- \mu^-) \}$.
The standard model (SM) predictions of the charge asymmetry induced by CP-violation are small in magnitude compared to the current
experimental precision, so non-zero measurements would indicate new sources of CP-violation.  D0 has previously published three
measurements of the CP-violating like-sign dimuon charge asymmetry in $p\overline{p}$ collisions at $\sqrt{s}$ = 1.96 TeV at the Fermi~lab Tevatron collider.  
These measurements were at integrated luminosities of 1 fb$^{-1}$ ~\cite{1 inv-fb}, 6.1 fb$^{-1}$ ~\cite{6 inv-fb}, and 9 fb$^{-1}$ ~\cite{9 inv-fb},
each observing $A_{CP}$ with differences from the standard model predictions of 1.7 to 3.9 $\sigma$ significance.  
This is one of only a few apparent inconsistencies with the standard model.

The major questions to be answered are whether these observations of deviation from the standard model are real, is our understanding of the SM
complete, and is there something else going on beyond the SM?

A new analysis of the full Run II data set of 10.4 fb$^{-1}$ with improved background subtraction and upgraded analysis methodology is currently under 
collaboration review.  It is not yet ready for public release, so I can only give a status report, show the checks performed using single inclusive muons, and
give an indication of the expected sensitivities.  The slides presented at DPF2013, with additional figures, are available at Reference \cite{talk}.

\section{Theoretical Framework}

In the standard model, one manifestation of CP-violation is in the mixing of the neutral $B$ mesons
$B^0 \leftrightarrow \overline{B}^0$ and $B_S^0 \leftrightarrow \overline{B}_S^0$.  
This asymmetry can be observed in the decays of pairs of particles containing $b$ and $\overline{b}$ quarks.  Pairs of $b$ and $\overline{b}$ quarks are produced symmetrically in the $p\overline{p}$ collisions.
These quarks hadronize into pairs of $B$ and $\overline{B}$ particles, including baryons.  
For example, particles containing $b$ quarks can have the decay chain $b \rightarrow \mu^- + X$,
while particles containing $\overline{b}$ quarks can have the decay chain $\overline{b} \rightarrow \mu^+ + X$.  For these direct decays, the  negative charge of this ``right-sign'' muon will 
tag the $b$ flavor of the parent quark.  The other $\overline{b}$ quark can decay $\overline{b} \rightarrow \mu^+ + X$, producing an opposite-sign dimuon $\mu^+ \mu^-$ pair.
However, for example, the parent $b$ quark could  hadronize into a $\overline{B}^0$ which could then oscillate into a $B^0$ which then can decay into a ``wrong-sign'' $\mu^+ + X$, schematically $b \rightarrow 
\overline{b} \rightarrow \mu^+$.  So the oscillation of either $B^0 \leftrightarrow \overline{B}^0$ or $B_S^0 \leftrightarrow \overline{B}_S^0$ can produce same sign dimuons.  CP-violation occurs 
if the rate $\Gamma (B^0 \rightarrow \overline{B}^0)$ does not equal the rate $ \Gamma(\overline{B}^0 \rightarrow B^0)$, 
or $\Gamma(B_S^0 \rightarrow \overline{B}_S^0) \ne \Gamma(\overline{B}_S^0 \rightarrow B_S^0)$,
which can produce an observable charge asymmetry for the like-sign dimuons for $\Gamma(\mu^+\mu^+) \ne  \Gamma(\mu^-\mu^-)$.
The sequential decays $b \rightarrow c \rightarrow \mu^+$ are background sources of wrong-sign muons.
Using the Pythia ~\cite{pythia} simulation for the total number of muons from $b$-particles, D0 observes approximately
73\% $b \rightarrow \mu^-$; 11\% $b \rightarrow \overline{b} \rightarrow \mu^+$; and 16\% $b \rightarrow c \rightarrow \mu^+$.

Previously, the only source of charge asymmetry for these like-sign dimuons in the standard model was considered to be via CP-violation in mixing.  
The predicted magnitude of this effect in D0 is
$A_{CP}^{\rm mixing}$(SM) = (-0.008 $\pm$ 0.001)\%.
Recently, however, G. Borissov and B. Hoeneisen ~\cite{interference} have calculated an additional CP-violating contribution due to the interference between processes involving
identical CP-definite states that can be reached both via mixing and non-mixing paths.  For example, a $B^0$ can produce a wrong sign $\mu^+$ via the CP-even final state $D^- D^+$ where
the $D^+ \rightarrow \mu^+ X$ decay.  The $B^0$ can produce $D^- D^+$ either directly or by first oscillating into $\overline{B}^0$ which can also decay into $D^- D^+$.  There is 
interference and CP-violation between the two paths.  
This interference does not contribute to $a_{CP}$ for single muons since the rates for $D^+ \rightarrow \mu^+$ and $D^- \rightarrow \mu^-$ balance. 

To set the scale, for the D0 like-sign dimuon charge asymmetry, this interference term is calculated to be
$A_{CP}^{\rm int}$(SM) = (-0.035 $\pm$ 0.008)\%, or about 4 times $ A_{CP}^{\rm mixing}$(SM).
So $A_{CP}(SM)$ = $A_{CP}^{\rm mixing}$(SM) + $A_{CP}^{\rm int}$(SM) = {-0.043 $\pm$ 0.010)\%
which can be compared with the previously measured $A_{CP}$(D0, 9 fb$^{-1}$) = (-0.276 $\pm$ 0.092)\% ~\cite{9 inv-fb}.

$A_{CP}^{\rm int}$ is linearly dependent on $\Delta\Gamma_d/\Gamma_d$, the
ratio of the difference of the widths of the light and heavy members of the mass eigenstates, $\Gamma(B_d^{\rm light})-\Gamma(B_d^{\rm heavy})$, to their average. 
This, then, gives the possibility of measuring $\Delta\Gamma_d/\Gamma_d$ in this analysis. 
The current World Average~\cite{HFAG-2012} for $\Delta\Gamma_d/\Gamma_d$ is (1.5 $\pm$ 1.8)\%, while the SM prediction~\cite{Lenz} is (0.42 $\pm$ 0.08)\%.  
It is anticipated that in this analysis, D0 will be able to measure $\Delta\Gamma_d/\Gamma_d$ to $\approx$ 1\% (absolute) precision.
Interference between such mixed and non-mixed paths for $B_s^0$ is too small to be observable in the D0 data set.
The analogous $\Delta\Gamma_s/\Gamma_s$ for $B_s^0$ is much smaller than for $B_d^0$ and is already well determined~\cite{HFAG-2012,Lenz}.

\section{Experimental Situation}

The three prior D0 analyses ~\cite{1 inv-fb,6 inv-fb,9 inv-fb} of the CP-violating like-sign dimuon asymmetry consistently measured $A_{CP}$ in the range -0.25 \% to -0.28 \%  
which differed from the predictions of the standard model (assuming mixing only, without the interference between the mixed and non-mixed paths) by significances of 1.7 to 3.9 $\sigma $.
Table~1 shows the evolution of the D0 measurement of $A_{CP}$ with increasing integrated Luminosity and sophistication of the analysis.  
The $A_{CP}$ results from this analysis of the full 10.4~fb$^{-1}$ data set, with reduced systematic uncertainties will be compared with the sum of $A_{CP}^{\rm mixing} + A_{CP}^{\rm int}$.

\begin{table}[t]
\begin{center}
\begin{tabular}{l|cccc}  
$\int {\cal L}$ $dt$ & Asymmetry $A_{CP}$ &  deviation from SM & Reference \\ \hline
 1.0 fb$^{-1}$  &  (-0.28 $\pm$ 0.13 $\pm$ 0.09)\%    &     1.7 $\sigma$      &    ~\cite{1 inv-fb} (2006) \\
 6.1 fb$^{-1}$  &  (-0.252 $\pm$ 0.088 $\pm$ 0.092)\%    &     3.2 $\sigma $     &    ~\cite{6 inv-fb} (2010) \\
 9.0 fb$^{-1}$  &  (-0.276 $\pm$ 0.067 $\pm$ 0.063)\%    &     3.9 $\sigma $     &    ~\cite{9 inv-fb} (2011) \\
10.4 fb$^{-1}$  &  (???? $\pm$ 0.064 $\pm$ 0.055)\%    &     ? $\sigma $     &    in preparation (2013)  \\ \hline
\end{tabular}
\caption{Evolution of the D0 measurement of $A_{CP}$ for the like-sign dimuon charge asymmetry.}
\label{tab:evolution}
\end{center}
\end{table}

\section{Experimental Methodology}

Why is D0 ~\cite{D0 detector} a good place to measure the like-sign dimuon charge asymmetry?
The CP-symmetric initial $p\overline{p}$ state does not have a charge asymmetry in the central region, or when integrated over a symmetric range of $\pm$ $\eta$.
Due to the large amount of hadronic absorption in the U-LAr Calorimeter and the tracking through the muon toroids ~\cite{D0 muon system}, 
D0 has excellent muon identification ~\cite{D0 muon reconstruction}.
The magnetic field directions in the central tracking solenoid magnet ~\cite{D0 detector} and in the muon toroids ~\cite{D0 muon system} are cycled through all combinations
on a regular basis which allows for cancellation of first-order effects due to instrumental asymmetries.

%Asymmetries for single inclusive muons and like-sign dimuons are denoted by lower case $a_{CP}$ and upper case $A_{CP}$ variables, respectively.
%We do not expect to observe charge asymmetries for the single inclusive muon data sample.  This data sample serves as a closure or consistency check that we are not generating
%false asymmetries through the apparatus, the acceptances, the analysis, or the background subtractions.

We observe a sample of 2.2 x $10^9$ single $\mu^\pm$, 2.2 x $10^7$ opposite sign $\mu^+ \mu^-$, and 6.2 x $10^6$ $\mu^\pm \mu^\pm$ like-sign dimuons. 

For analysis, the data are primarily divided into three muon (transverse) Impact Parameter (IP) bins (IP=1, 2, 3) corresponding to (0-50 $\mu m$), (50-120 $\mu m$), and (120-3000 $\mu m$),
and a sum integrating over ALL IP.
The muons of interest from $b$ decays are predominantly at large IP, while the muons from the decays of kaons and pions are predominantly at small IP, 
since the parent kaons and pions have already been tracked from the primary vertex before decaying. 
Each of these four IP sets are also sub-divided into a combination of nine ($p_\top,|\eta|$) bins:

Bins \# 1-3:  \space\space 0 $\le |\eta| \le$ 0.7 , $p_\top$ = 4.2-5.6, 5.6-7, 7-25 GeV; 

Bins \# 4-5: 0.7 $\le |\eta| \le$ 1.2 , $p_\top$ = 3.5-5.6, 5.6-25 GeV;

Bins \# 6-9: 1.2 $\le |\eta| \le$ 2.2 , $p_\top$ = 1.5-3.5, 3.5-4.2, 4.2-5.6, 5.6-25 GeV.

Standard D0 single- and multi-muon triggers and analyses ~\cite{1 inv-fb,6 inv-fb,9 inv-fb} are used, along with slightly tighter requirements on tracking quality.
To ensure that the muon candidates penetrate through the muon toroids, we require either $p_\top > $ 4.2 GeV or $|p_z| >$ 5.2 GeV.  
We also require $p_\top < $ 25 GeV to avoid muons from $W^\pm$ and $Z^0$ decays.  The dimuon invariant mass $M_{\mu\mu}$ is required to be greater than 2.8 GeV to avoid both muons from the decay chain of the same $b$ quark  
$b \rightarrow \mu^- $ $\nu$  $c$ \space $(\rightarrow \mu^+)$.

Based on the muon charge configuration of number of events observed $n^\pm$, $N^{++}$, and $N^{--}$, 
the raw (observed) asymmetries (in each IP, $p_\top, |\eta|$) bin are defined as:
$A = (N^{++}-N^{--})$/$(N^{++}+N^{--})$ and $a = (n^+ - n^-)$/$(n^+ + n^-)$ for like-sign dimuons and for single inclusive muons, respectively.
The background subtracted residual CP asymmetries are
$a_{CP} = a - a_{bkg}$ and $A_{CP} = A - A_{bkg}$, where $a_{bkg} = a_\mu + f_K  a_K + f_\pi  a_\pi + a_p $.
$f_K$ is the fraction of charged kaons in the $\mu$ sample, measured using dedicated channels with final-state kaons reconstructed as muons.
$a_K$ is the asymmetry is due to the difference in the inelastic cross sections between $K^+$ and $K^-$.
$f_K  a_K$ is typically +0.62\% and is the dominant background term at low IP.
$a_\mu$ is the muon detector charge asymmetry measured with $J/\psi \rightarrow \mu^+ \mu^-$.
$a_\mu$ is typically -0.29\% and is the next dominant background term.  $f_\pi  a_\pi$ and $a_p$ are considerably smaller.
In the current analysis, $f_K$ and $f_\pi$ are cross-checked using tracks measured in both the central tracker and in the local muon detector trackers.
The differences in the two measurements of the muon fractions are included in the systematic uncertainties. 

% condensing this section
%$f_K$ is the fraction of charged kaons in the $\mu$ sample, measured from reconstructed kaons.
%$K^{*0} \rightarrow \pi^- + K^+ (\rightarrow \mu^+)$ and $K^{*+} \rightarrow \pi^+ K_S^0 (\rightarrow \pi^+ \pi^-)$
%and then converting $f_{K^{*0}} + f_{K^{*+}} \rightarrow f_{K^\pm}$ by isospin invariance.  
%$a_K$ is the asymmetry is due to the difference in the inelastic cross sections for K^+ and K^0.
%#$\sigma_{inelastic}(K^-)$ being greater than $\sigma_{inelastic}(K^+)$ and is measured with	$K^{*0} \rightarrow \pi^- K^+ (\rightarrow \mu^+ \nu)$ and its charge conjugate and $\phi \rightarrow K^+ K^-$ (with $K^\pm \rightarrow \mu^\pm \nu$).
%$f_K  a_K$ is typically +0.62\% and is the dominant background term at low IP.
% condense here too
%$a_\mu$ is the muon detector charge asymmetry measured with $J/\psi \rightarrow \mu^+ \mu^-$ without a muon trigger, 
%but identified and measured just using the central tracker.
%$a_\mu$ is typically -0.29\% and is the next dominant background term.  $f_\pi  a_\pi$ and $a_p$ are considerably smaller.

\section{Checks and Projections of Sensitivities}

The standard model predicts the magnitudes of the CP asymmetries $a_{CP}$ for single inclusive muons for all of the (IP, $p_\top, |\eta|$) bins which are well 
below the sensitivity limits of the D0 10.4 fb$^{-1}$ data.  
Therefore, D0 expects the measurements of these CP asymmetries for single inclusive muons to be consistent with zero.    
The single muon data serves as a closure test or consistency check that we are not generating false asymmetries through the apparatus, 
the acceptances, the analysis, or the background subtractions. To illustrate the preliminary resolutions and data scatter for the 10.4 fb$^{-1}$ data, 
Figure~1 shows the raw, observed asymmetries $a$ (upper histogram), measured background asymmetries $a_{bkg}$ (upper data points), and the background subtracted CP violating
asymmetries $a_{CP} = a - a_{bkg}$ (lower data points) for each of the 9 ($p_\top, |\eta|$) bins, for ALL IP and for the three IP bins. 
The $a_{CP} = a - a_{bkg}$ plot for ALL IP demonstrates consistency with the expected zero asymmetry, along with the uncertainty spread for the average over the nine ($p_\top, |\eta|$) bins.

%% - remove this figure 19sept2013 - re-install 20sept2013
\begin{figure}[htb]
\centering
\includegraphics[height=5.00in]{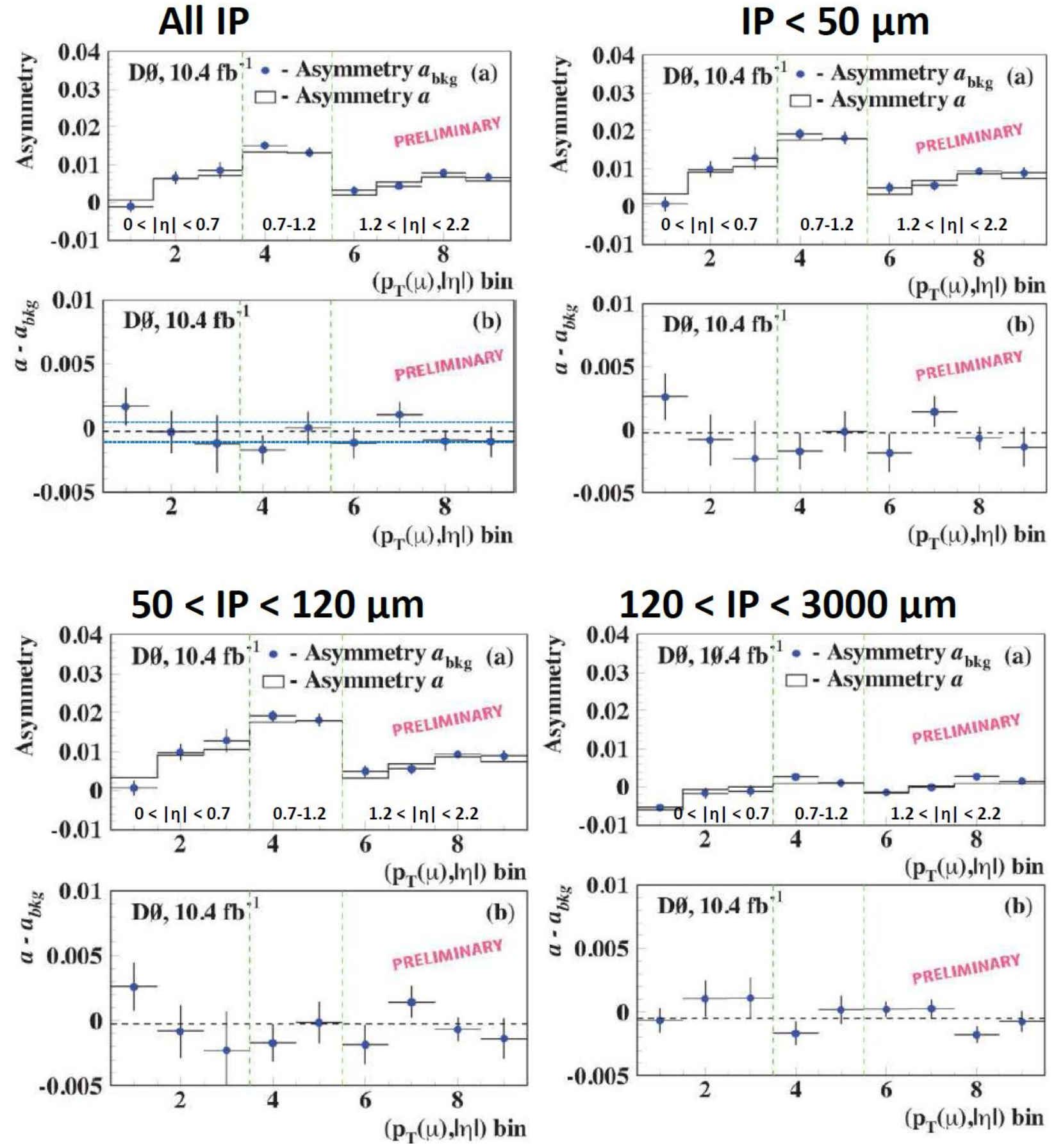}
\caption{The raw, background, and CP violating asymmetries for single inclusive muons for ALL IP and the three IP ranges and the 9 bins in $(p_\top,|eta|)$.}
\label{fig:aCP}
\end{figure}
%%%

Before the interference term was included in the like-sign dimuon phenomenology, 
the 9 fb$^{-1}$ analysis ~\cite{9 inv-fb} used only two IP bins of IP $<$ 120 $\mu m$ and IP $>$ 120 $\mu m$ and the sum over all IP.  
This produced the three linear correlation bands between $a_{sl}^d$ and $a_{sl}^s$ along with their correlated 68\% and 95\% CL uncertainty ellipses in Figure~2. 
Given the fitting over three independent IP bins and the 9 ($p_\top, |\eta|$) bins, and the upgraded background subtraction and analysis, 
D0 anticipates that the $areas$ of these uncertainty ellipses will decrease by $\approx$ 44 \%.
The D0 direct measurements of $a_{sl}^d$ = (0.68 $\pm$ 0.47)\% from $B^0 \rightarrow D^{(*)-} \mu^+ X$ ~\cite{asld} 
and $a_{sl}^s$ = (-1.12 $\pm$ 0.76)\% from $B_s^0 \rightarrow D_s^- \mu^+ X$ ~\cite{asls} are overlayed for comparison.

%%% - replace by single figure 19sept2013
\begin{figure}[htb]
\centering
\includegraphics[height=3.90in]{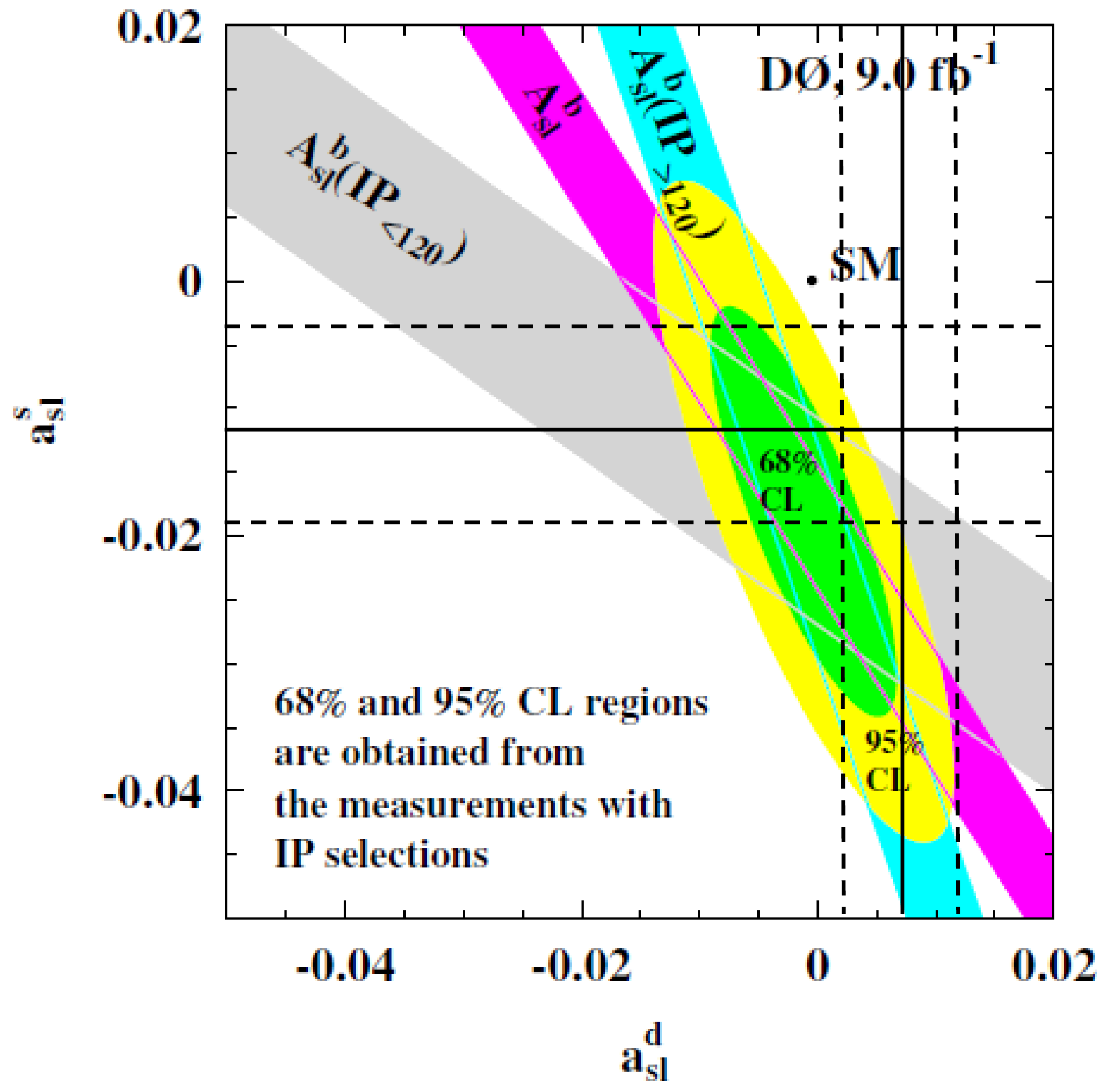}
\caption{The $a_{sl}^d$ vs. $a_{sl}^s$ measurement and uncertainty ellipse contours for the prior 9 fb$^{-1}$ D0 same sign dimuon charge asymmetry analysis ~\cite{9 inv-fb} 
compared to the SM predictions.   
The horizontal and vertical solid and dashed lines represent the measurements $\pm$ 1 $\sigma$ bands 
from recent D0 inclusive semi-leptonic decay measurements of $a_{sl}^d$ and $a_{sl}^s$~\cite{asld,asls}.  }
\label{fig:aslq_overlay}
\end{figure}
%%%

\section{Summary}

D0 is preparing the final release of the analysis of the CP violating like-sign dimuon analysis based on the full 10.4 fb$^{-1}$ Run II data set.  This result is anticipated to
have an uncertainty on the asymmetry $A_{CP}$ of $\pm$ 0.084 \%  (stat. + syst.), allowing more stringent comparison with the predictions of the standard model.
This analysis will also decrease the area of the uncertainty ellipse for the semi-leptonic decay asymmetries $a_{sl}^d$ and $a_{sl}^s$ by a factor of $\approx$ 44 \%.  

Remaining questions to be addressed are whether the entire like-sign dimuon charge asymmetry could be due to a large value of $\Delta \Gamma_d/\Gamma_d$; whether there are
still missing SM contributions not included in the calculation of $A_{CP}$; and whether the D0 observation of significant deviations from the SM predictions is real. 
Addressing the latter will require verification by other experiments. 

The final D0 paper on CP-violation for like-sign dimuons based on this analysis has been submitted for publication in early October, 2013 \cite{arXiv}.

\Acknowledgments
We thank the staffs at Fermilab and collaborating institutions,
and acknowledge support from the DOE and NSF
(USA); CEA and CNRS/IN2P3 (France); FASI, Rosatom
and RFBR (Russia); CNPq, FAPERJ, FAPESP and
FUNDUNESP (Brazil); DAE and DST (India);
Colciencias (Colombia); CONACyT (Mexico); KRF and
KOSEF (Korea); CONICET and UBACyT (Argentina);
FOM (The Netherlands); STFC and the Royal Society
(United Kingdom); MSMT and GACR (Czech Republic);
CRC Program and NSERC (Canada); BMBF and DFG
(Germany); SFI (Ireland); The Swedish Research Council
(Sweden); and CAS and CNSF (China).

I also would like to thank the members of the D0 Collaboration and especially 
Guennadi Borissov (Lancaster University) and 
Bruce Hoeneisen (Universidad San Francisco de Quito) 
for their discussions, guidance, and sharing materials.

\end{document}